%% file: letter.tex
\documentclass[aps,prl,twocolumn,superscriptaddress,nofootinbib,nobibnotes]{revtex4-2}

\input{packages}

\def \be {\begin{equation}}
\def \ee {\end{equation}}

\newcommand{\lcdm}{$\mathrm{\Lambda CDM}$\xspace}
\newcommand{\Hc}{\ensuremath{H_0}\xspace}

\newcommand{\Om}{\ensuremath{\Omega_{\mathrm m}}\xspace}

\newcommand{\kext}{\ensuremath{\kappa_{\rm ext}}\xspace}

\newcommand{\lint}{\lambda_{\rm int}}

\newcommand{\Jtwelve}{SDSS\ J1206$+$4332\xspace}

\newcommand{\HEeleven}{HE\ 1104$-$1805\xspace}

\def\ksmpc{${\, \mathrm{km}\, \mathrm{s}^{-1}\, \mathrm{Mpc}^{-1}}$\xspace}

\newcommand{\fref}[1]{Fig.~\ref{#1}}

\newcommand{\hnotfiducial}{71.1^{+2.4}_{-2.3}}

\newcommand{\nslacstot}{15}
\newcommand{\nsltwossel}{8}

\newcommand{\glee}{\textsc{glee}\xspace}
\newcommand{\lenstro}{\textsc{lenstronomy}\xspace}

\begin{document}

\title{TDCOSMO 2026: new insights on the Hubble tension from strong lensing time delays}

\author{TDCOSMO Collaboration}
\noaffiliation

\author{S.~Birrer}
\affiliation{Department of Physics and Astronomy, Stony Brook University, Stony Brook, NY 11794, USA}

\author{M.~Cappellari}
\affiliation{Sub-Department of Astrophysics, Department of Physics, University of Oxford, Denys Wilkinson Building, Keble Road, Oxford OX1 3RH, UK}

\author{F.~Courbin}
\affiliation{ICC-UB Institut de Ci\`encies del Cosmos, Universitat de Barcelona, Mart\'i Franqu\`es, 1, 08028 Barcelona, Spain}
\affiliation{Institut Catal\`a de Recerca i Estudis Avan\c{c}ats (ICREA), Pg. Llu\'is Companys 23, 08010 Barcelona, Spain}
\affiliation{Institut d'Estudis Espacials de Catalunya (IEEC), Edifici RDIT, Campus UPC, Castelldefels, 08860 Barcelona, Spain}

\author{C.~D.~Fassnacht}
\affiliation{Department of Physics and Astronomy, University of California, Davis, 1 Shields Ave., Davis, CA 95616, USA}

\author{P.~Fleury}
\affiliation{Laboratoire Univers et Particules de Montpellier (LUPM), CNRS \& Universit\'{e} de Montpellier (UMR-5299), Parvis Alexander Grothendieck, F-34095 Montpellier Cedex 05, France}

\author{J.~A.~Frieman}
\affiliation{SLAC National Accelerator Laboratory, Menlo Park, CA 94025, USA}

\author{A.~Galan}
\email[Corresponding author: ]{aymeric.galan@gmail.com}
\affiliation{Department of Astronomy, University of Geneva, ch. d'Ecogia 16, 1290 Versoix, Switzerland}

\author{X.-Y.~Huang}
\affiliation{Department of Physics and Astronomy, Stony Brook University, Stony Brook, NY 11794, USA}

\author{D.~P.~Johnson}
\affiliation{Laboratoire Univers et Particules de Montpellier (LUPM), CNRS \& Universit\'{e} de Montpellier (UMR-5299), Parvis Alexander Grothendieck, F-34095 Montpellier Cedex 05, France}

\author{S.~Khanolkar}
\affiliation{Department of Physics and Astronomy, University of California, Los Angeles, CA 90095, USA}

\author{S.~Knabel}
\affiliation{Department of Physics and Astronomy, University of California, Los Angeles, CA 90095, USA}

\author{H.~Krizic}
\affiliation{D\'epartement de Physique Th\'eorique, Universit\'e de Gen\`eve, 24 quai Ernest-Ansermet, CH-1211 Gen\`eve 4, Switzerland}

\author{M.~Millon}
\email[Corresponding author: ]{martin.millon@unige.ch}
\affiliation{D\'epartement de Physique Th\'eorique, Universit\'e de Gen\`eve, 24 quai Ernest-Ansermet, CH-1211 Gen\`eve 4, Switzerland}

\author{T.~Morishita}
\affiliation{Astronomical Institute, Graduate School of Science, Tohoku University, 6–3 Aoba, Sendai 980-8578, Japan}

\author{V.~Motta}
\affiliation{Instituto de F\'{\i}sica y Astronom\'{\i}a, Universidad de Valpara\'{\i}so, Avda. Gran Breta\~na 1111, Valpara\'{\i}so, Chile}

\author{P.~Mozumdar}
\email[Corresponding author: ]{pmozumdar@astro.ucla.edu}
\affiliation{Department of Physics and Astronomy, University of California, Los Angeles, CA 90095, USA}

\author{E.~Paic}
\affiliation{Research Center for the Early Universe, Graduate School of Science, The University of Tokyo, 7-3-1 Hongo, Bunkyo-ku, Tokyo 113-0033, Japan}

\author{H.~Paugnat}
\affiliation{Department of Physics and Astronomy, University of California, Los Angeles, CA 90095, USA}

\author{A.~G.~Schweinfurth}
\affiliation{Technical University of Munich, TUM School of Natural Sciences, Department of Physics, James-Franck-Str.~1, 85748 Garching, Germany}
\affiliation{Max-Planck-Institut für Astrophysik, Karl-Schwarzschild-Str.~1, 85748 Garching, Germany}

\author{A.~J.~Shajib}
\affiliation{Department of Astronomy \& Astrophysics, University of Chicago, Chicago, IL 60637, USA}
\affiliation{Kavli Institute for Cosmological Physics, University of Chicago, Chicago, IL 60637, USA}
\affiliation{Center for Astronomy, Space Science and Astrophysics, Independent University, Bangladesh, Dhaka 1229, Bangladesh}

\author{W.~Sheu}
\affiliation{Department of Physics and Astronomy, University of California, Los Angeles, CA 90095, USA}

\author{D.~Sluse}
\affiliation{STAR Institute, University of Li{\`e}ge, Quartier Agora, All\'ee du six Ao\^ut 19c, 4000 Li\`ege, Belgium}

\author{M.~Stiavelli}
\affiliation{Space Telescope Science Institute, 3700 San Martin Dr., Baltimore, MD 21218, USA}

\author{S.~H.~Suyu} 
\affiliation{Technical University of Munich, TUM School of Natural Sciences, Department of Physics, James-Franck-Str.~1, 85748 Garching, Germany}
\affiliation{Max-Planck-Institut für Astrophysik, Karl-Schwarzschild-Str.~1, 85748 Garching, Germany}

\author{T.~Treu}
\affiliation{Department of Physics and Astronomy, University of California, Los Angeles, CA 90095, USA}

\author{F.~Urcelay}
\affiliation{Sub-Department of Astrophysics, Department of Physics, University of Oxford, Denys Wilkinson Building, Keble Road, Oxford OX1 3RH, UK}

\author{D.~M.~Williams}
\affiliation{Department of Physics and Astronomy, University of California, Los Angeles, CA 90095, USA}

\author{K.~C.~Wong}
\affiliation{Research Center for the Early Universe, Graduate School of Science, The University of Tokyo, 7-3-1 Hongo, Bunkyo-ku, Tokyo 113-0033, Japan}

\begin{abstract}
We discuss the measurement of the Hubble constant ($H_0$) in flat $\Lambda$ cold dark matter cosmology based on the blind hierarchical analysis of 11 multiply-imaged quasars (hereafter the TDCOSMO-2026 sample) and described in greater detail in a companion paper. In addition to the time delays, we use lens models based on high-resolution imaging from the Hubble Space Telescope (HST) and from the James Webb Space Telescope~(JWST), analyzed jointly with axisymmetric dynamical models based on JWST-NIRSpec, VLT-MUSE, and Keck-KCWI spectroscopy. Our primary result $H_0=\hnotfiducial$\ksmpc\ is based on the TDCOSMO-2026 sample, in combination with $\Omega_{\rm m}$ constraints from the DES-Dovekie supernova (SN) dataset. The result is robust to variations in analysis methods, priors, and external datasets. Our 3\% precision accounts for the mass-sheet degeneracy. Importantly, our measurement is independent of the local distance ladder (LDL). Among individual late-universe probes, only the traditional LDL is more precise. We discuss our result in the context of the Hubble tension, the difference between $H_0$ measured in the late universe and the prediction based on early-universe probes such as the cosmic microwave background (CMB). We construct combinations of independent late-universe probes and show that the tension with the \textit{Planck} CMB prediction is robust to excluding specific measurements or methods. For example, we construct a ``no-SN'' $H_0$ measurement, by replacing the DES-Dovekie constraint on $\Omega_{\rm m}$ with that from DESI baryon acoustic oscillations, and then combining TDCOSMO-2026 with independent non-SN late-universe probes. This ``no-SN'' late-universe probe is in 3.7$\sigma$ tension with \textit{Planck}.  Even after completely excluding the LDL, the tension with \textit{Planck} persists at the 2.7$\sigma$ level. 
We conclude by describing the path to 1.5\% precision with strong lensing time delays.
\end{abstract}

\maketitle

\section{Introduction}

The Hubble constant ($H_0$) is the current expansion rate of the universe. It has now been measured to better than 1\% precision using the relationship between redshift and distance known as the Hubble--Lema\^itre law \citep{H0DN:2026}. Given a cosmological model, $H_0$ can also be predicted from measurements of the properties of the Universe at earlier times. For example, in a flat \lcdm cosmology, observations of the anisotropies of the cosmic microwave background can be used to predict $H_0$ with precision better than 1\% \citep{Planck2020}. The direct (or late-universe) measurement by \citet{H0DN:2026} and the predicted value (or early-universe) by \textit{Planck} \citep{Planck2020} differ by $\sim$8--9\%. This statistically significant difference is known as the Hubble Tension. The tension is not restricted to the two measurements listed above: early universe probes generally are in tension with a variety of late universe probes \citep{cosmoverse25}. If the tension is real, and not due to systematic errors in multiple measurements, it implies that \lcdm is not a correct description of the Universe. At a theoretical level, there is no consensus about what modifications of the standard cosmological model could solve the tension \cite[see][and references therein]{cosmoverse25}. From an observational point of view, the only way to ensure that the tension is not driven by systematic errors is to develop multiple independent methods. Each of these methods should have sufficient precision to confirm or reject the tension. 

Strong gravitational lensing time-delay cosmography \citep{TreuMarshall2016} is a method to measure $H_0$ that offers many advantages. It is independent of all other methods. It measures absolute angular-diameter distances directly, without the need for external calibrators, dust correction, or photometric calibrations. It measures distances of several Gpc, well into the Hubble flow, and is thus immune to any local inhomogeneity \citep{Dalang2023}. The method has come a long way since it was first proposed by \citet{Refsdal1964} thanks to many improvements in data quality and analysis methods \citep[see, e.g.,][and references therein]{TreuMarshall2016,TSM22,Birrer:2024}.

In this Letter, we present the measurement of $H_0$ in flat \lcdm based on 11 strongly lensed quasars with measured time delays, reaching 3\% precision, a marked improvement compared to the 5\% precision reached in our previous 2025 milestone paper. This work is the culmination of many years of work by the TDCOSMO collaboration \citep{Millon:2020}, including papers \citep{Paic2026,Knabel2026,Mozumdar2026,sheu2026,GalanSchweinfurth:2026,Johnson2026} that present the most recent developments with respect to our 2025 measurement \citep{TDCOSMO2025} and feed into the analysis presented here. 
Following standard practice by our collaboration, our analysis is blinded with respect to cosmological parameters. Unblinding took place during a collaboration telecon on 23 Sep 2026, following internal review. The unblinded results are published without modification. A companion paper \citep[hereafter TDC26;][]{TDCOSMO2026_paper} presents the full details of the hierarchical analysis, the results in a variety of cosmologies, and a discussion of systematic errors. The aim of this letter is to present the \lcdm result in a short and accessible way to a broad audience of physicists and astronomers.

\section{Summary of the method}
\label{sec:method}

We give here a brief summary of time-delay cosmography, referring the reader to the companion paper TDC26 and extensive reviews for further details \cite{TreuMarshall2016,TSM22,Birrer:2024}.

Conceptually, gravitational lensing is a non-linear mapping between the apparent position on the sky of a source (image plane) and the position it would have without a deflector in front (source plane). Under appropriate circumstances -- if the deflector is sufficiently dense and the source and deflector are sufficiently well aligned -- multiple images appear to the observer. This phenomenon is called strong lensing. 

In order to transform this mapping of angular quantities into physical distances, one needs dimensional quantities. Fortunately, multiple images do not appear to the observer at the same time. Similarly to Fermat's principle in standard optics, Fermat's principle in gravitational optics states that images form at the extrema of the arrival time surface \citep{Schneider1985,Blandford1986}. The difference in arrival time between the images is known as the time delay. When it is measured, it provides an absolute scale to the problem, and thus can be used to convert angular quantities to angular-diameter distances. In turn, angular-diameter distances can be used to reconstruct the expansion history of the universe and therefore measure $H_0$ and other cosmological parameters.

In mathematical terms, the time delay between two images at positions $\boldsymbol{\theta}_{\rm A}$ and $\boldsymbol{\theta}_{\rm B}$, originating from a source at $\boldsymbol{\beta}$, is given by
\begin{equation}\label{eqn:time_delay}
\Delta t_{\rm AB} = \frac{D_{\Delta t}}{c} \left[\tau(\boldsymbol{\theta}_{\rm A}, \boldsymbol{\beta}) - \tau(\boldsymbol{\theta}_{\rm B}, \boldsymbol{\beta}) \right],
\end{equation}
where the Fermat potential is
\begin{equation}\label{eqn:fermat_potential}
\tau(\boldsymbol{\theta}, \boldsymbol{\beta}) = \frac{1}{2}(\boldsymbol{\theta} - \boldsymbol{\beta})^2 - \psi(\boldsymbol{\theta}),
\end{equation}
with $\psi(\boldsymbol{\theta})$ the two-dimensional projected gravitational potential, also known as the lensing potential. 

The time-delay distance \citep{Suyu2010} is defined as
\begin{equation} \label{eqn:ddt_definition}
D_{\Delta t} = (1 + z_{\rm d}) \frac{D_{\rm d} D_{\rm s}}{D_{\rm ds}},
\end{equation}
where $z_{\rm d}$ is the deflector redshift and $D_{\rm d}$, $D_{\rm s}$, and $D_{\rm ds}$ are angular-diameter distances, from the observer to the deflector, from the observer to the source, and from the deflector to the source, respectively. Like any distance, the time-delay distance is inversely proportional to $H_0$. Thus, a measurement of the time delay between two images, combined with constraints on the Fermat potential difference (e.g., from lensing and dynamics), yields $H_0$.

An important complication is the so-called mass-sheet degeneracy \citep[MSD;][]{Falco1985,Schneider2013} which affects lensing-only constraints. An infinite sheet of mass uniformly magnifies or demagnifies the image plane. Thus, if the source's absolute position or luminosity are not known, one can rescale the distances involved, to obtain exactly the same apparent configuration, leading to a mathematical degeneracy between the mass sheet and the inferred $H_0$. 

In our approach, we break the MSD by measuring the motion of the stars (line-of-sight velocity and velocity dispersion) within the deflector galaxy and by quantifying the effects of over and underdensities along the line of sight. In practice, this is quantified via two parameters: $\lint$ that quantifies the component of the mass sheet in the plane of the lens \citep{TDCOSMOIV} and the external convergence $\kext$ that quantifies the over- and under-density of a line of sight with respect to the background cosmology \citep{Johnson2026}.

With these definitions, the fundamental equation is then
\begin{equation}\label{eqn:time_delay_generalized}
    \Delta t_{\rm AB, \lambda} = (1-\kappa_{\rm ext}) \lambda_{\rm int} \frac{D_{\Delta t}}{c} \Delta \tau_{\rm AB},
\end{equation}
where the Fermat potential $\Delta \tau_{\rm AB}$ is the Fermat potential inferred without the inclusion of the internal and external mass sheet.  Since we are using stellar kinematics to break the MSD, the cosmological information comes from a combination of the time-delay distance and the angular-diameter distance to the deflector \cite{Birrer2016}.  As explained in detail in our companion paper, we carry out a hierarchical analysis: we fit eleven lenses with lens-specific parameters (e.g. $\Delta \tau$ and $\kext$ are specific to each lens) and use hyper-parameters to describe population level quantities such as cosmological parameters and the mean and scatter of $\lint$.

\section{Summary of the data}
\label{sec:data}

We briefly summarize the data used in this analysis, referring to the companion article for more details.

\begin{itemize}
    \item {\bf Time delays} are measured from monitoring campaigns utilizing a variety of 1--4 m ground-based telescopes over the past two decades, in particular the COSMOGRAIL program \citep{Millon2020, Courbin2005}. A recent summary of state-of-the-art techniques is given by \citet{Dux:2025}. 
    \item {\bf Lens models} are based on imaging data obtained with the Hubble (8 systems) and James Webb (3 systems) Space Telescopes, using a variety of cameras and filters. The most recent examples are discussed by \citep{sheu2026,GalanSchweinfurth:2026,Williams2026}. The lens modeling software packages \glee \citep{SuyuHalkola2010, Suyu2012b} and \lenstro \citep{Birrer2021lenstronomy} are used. Comparisons between software packages \citep{Shajib2022,GalanSchweinfurth:2026}, and between independent data sets \citep{Williams2026}, show that the results are robust.
    \item {\bf Redshifts of source and deflector:} are obtained from ground- and space-based spectroscopy. 
    \item {\bf Stellar kinematics} is derived from high signal-to-noise ratio spectroscopy. For six of the time-delay systems, integral field spectroscopy is available from JWST NIRSpec \citep{Shajib25b,Mozumdar2026,Knabel2026}. Spatially resolved kinematics from KCWI on Keck is available for two time-delay systems \citep{Shajib2023,sheu2026}. Spatially resolved kinematics from MUSE on VLT is available for two time-delay systems \citep{Paic2026,GalanSchweinfurth:2026}. For two time-delay systems,  integrated stellar velocity dispersion from MUSE on VLT is available \citep{TDCOSMO2025}. For the non-time-delay systems, spatially resolved kinematics from KCWI on Keck \citep{Knabel24} and integrated stellar velocity dispersions from a number of ground-based telescopes \citep{Mozumdar2025} are available.  
    \item {\bf External convergence} is estimated by comparing number counts in the fields with those found in Euclid Flagship Simulations \citep{Johnson2026}.
\end{itemize}

\begin{figure*}
    \centering
    \includegraphics[width=0.99\textwidth]{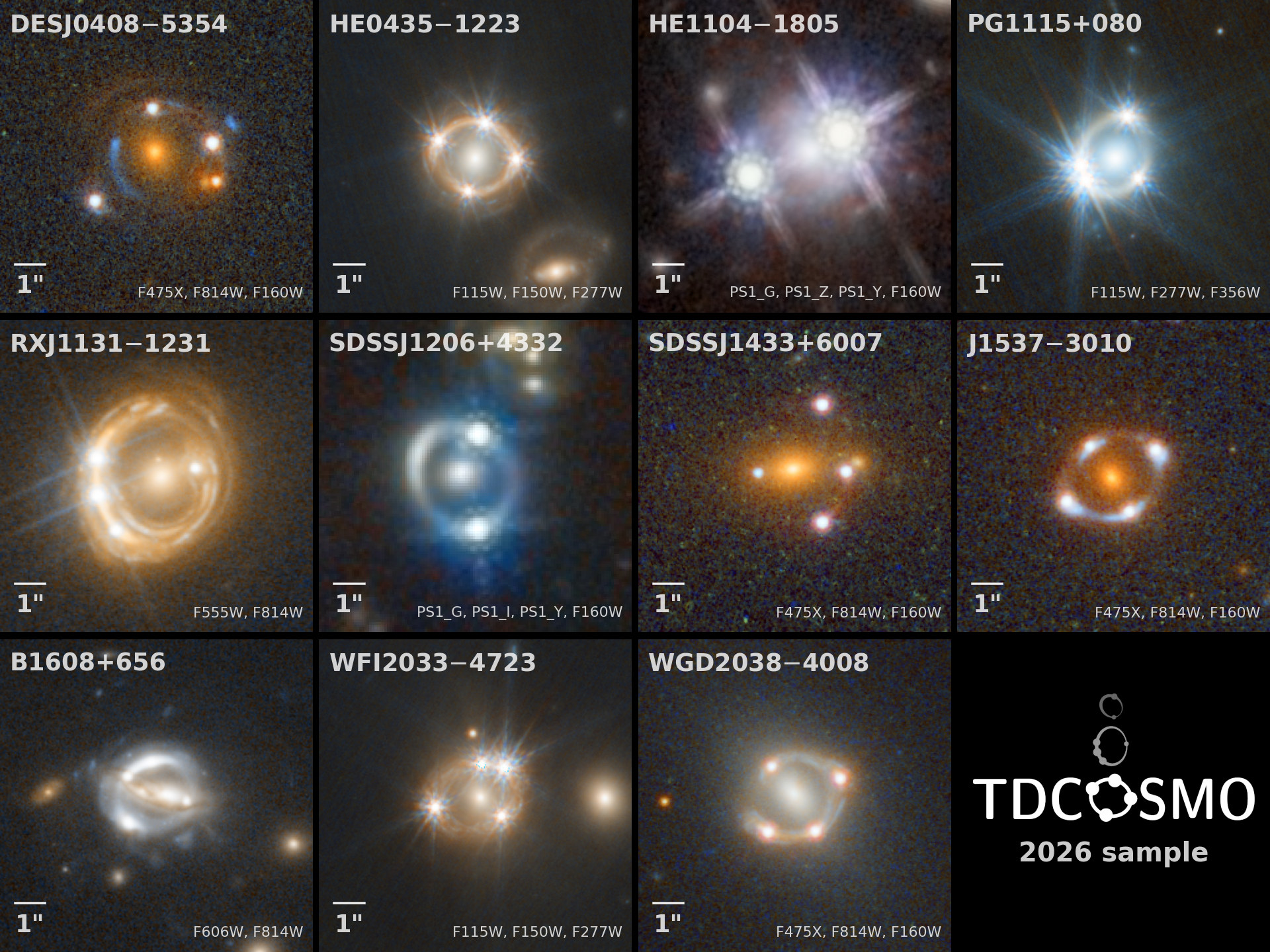}
    \caption{Montage of all the time-delay lenses in the TDCOSMO-2026 sample. All cutouts cover the same angular size and are aligned North up, East left. The color composite images are made from various instruments and filters from HST (F475X, F555W, F606W, F814W, F160W) and JWST/NIRCam (F115W, F150W, F277W). For \Jtwelve and \HEeleven, only one HST filter with suitable depth was available so we combined it with Pan-STARRS1 images \citep{PanstarrsData_Flewelling2020} to provide color information (PS1\_G, PS1\_I, PS1\_Z, PS1\_Y). We refer to the companion paper and to the individual papers for details about what bands and filters were used for lens modeling.}
    \label{fig:montage}
\end{figure*}

\section{Results}
\label{sec:results}

\subsection{The Hubble constant in flat $\Lambda$CDM from TDCOSMO}

\begin{figure*}
    \includegraphics[width=\textwidth]{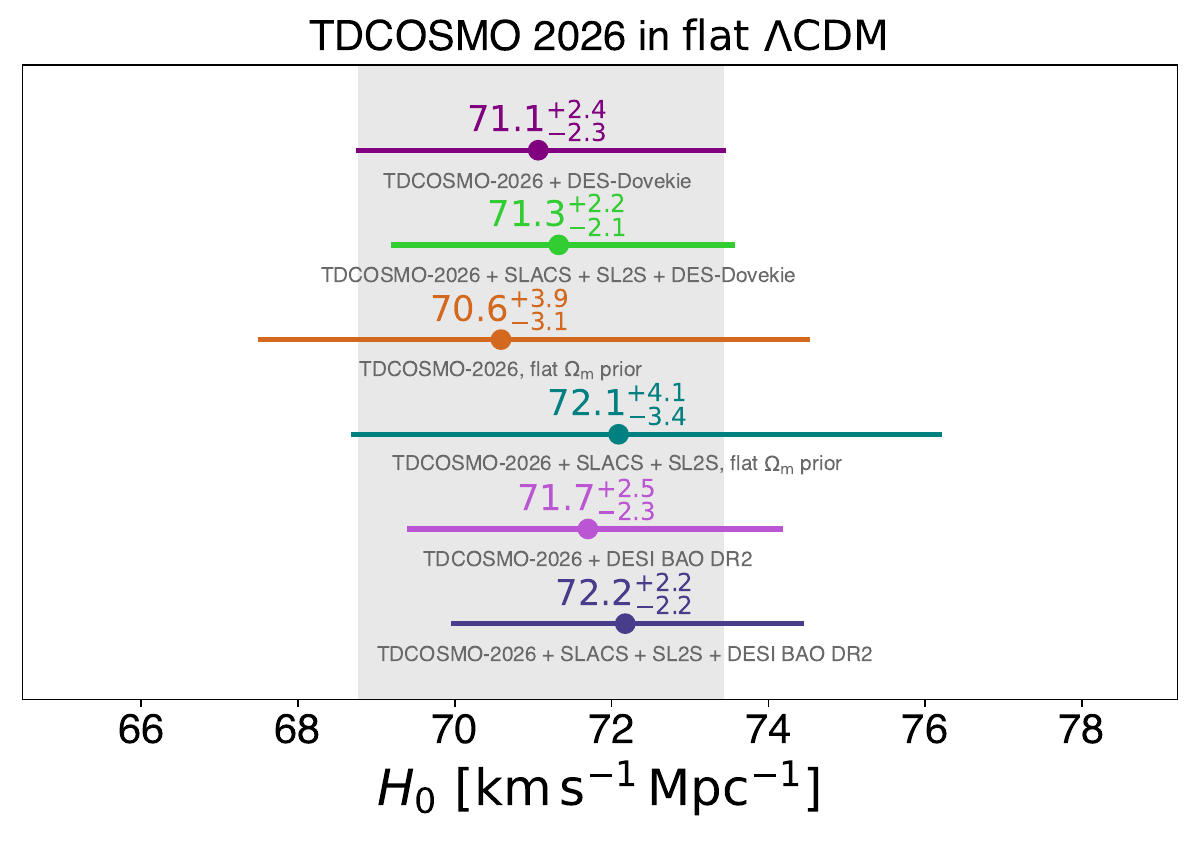}
    \caption{\Hc in flat \lcdm from the TDCOSMO-2026 time-delay lenses alone and combined with \nslacstot ~SLACS and \nsltwossel ~SL2S non-time-delay lenses. We show the impact of the \Om prior, either uniform, \Om $\in [0.05, 0.5]$, or constrained by the DES-Dovekie supernovae or DESI-DR2 BAO data. The gray band marks the 68\% interval of our baseline result, TDCOSMO-2026 with DES-Dovekie. The external datasets constrain only \Om and carry no information on the absolute distance scale: this is set solely by the time-delay lenses.}\label{fig:TDCOSMO}
\end{figure*}

Our baseline result is the measurement of $H_0$ based on the 11 time-delay lenses (Figure~\ref{fig:montage}), with an informative prior on $\Omega_{\rm m}$ taken from the DES-Dovekie supernova sample \citep{DESDovekie}. The baseline was decided prior to unblinding. Several variations of the analysis have been carried out, showing that the result is robust to details of the choice.  The examples in Figure~\ref{fig:TDCOSMO} include: changing the $\Omega_{\rm m}$  prior; adding information from the non time delay lens samples SLACS \citep{Bolton08} and SL2S \citep{Gavazzi12}, for which we have obtained high quality kinematics \citep{Knabel24,Mozumdar2025} to help constrain the internal mass-sheet parameter $\lint$ (TDCOSMO-2026+SLACS+SL2S). Other variations are discussed in the companion paper. In all cases, the result changes by less than the uncertainties. Extensive work on systematic errors, summarized in the companion paper and by \citet{TDCOSMO2025} shows that they are at the percent level, i.e., below statistical errors.

For simplicity, we continue our discussion using the baseline result. We use the TDCOSMO-2026 one based on $\Omega_{\rm m}$ prior informed by Baryon Acoustic Oscillations \citep[BAO][]{DESIDR2}, when we need to construct a measurement of $H_0$ that is independent of supernovae. Note that this prior does not rely on assuming the scale of the sound horizon at recombination. We also include in our discussion the most precise measurement based on TDCOSMO-2026+SLACS+SL2S.

\subsection{The Hubble tension}

The precision of TDCOSMO-2026 is not yet sufficient to settle the debate on the Hubble tension by itself. However, it is instructive to examine the significance of the tension by combining TDCOSMO-2026 with other independent late-universe probes. 

\begin{figure*}
    \includegraphics[width=0.9\textwidth]{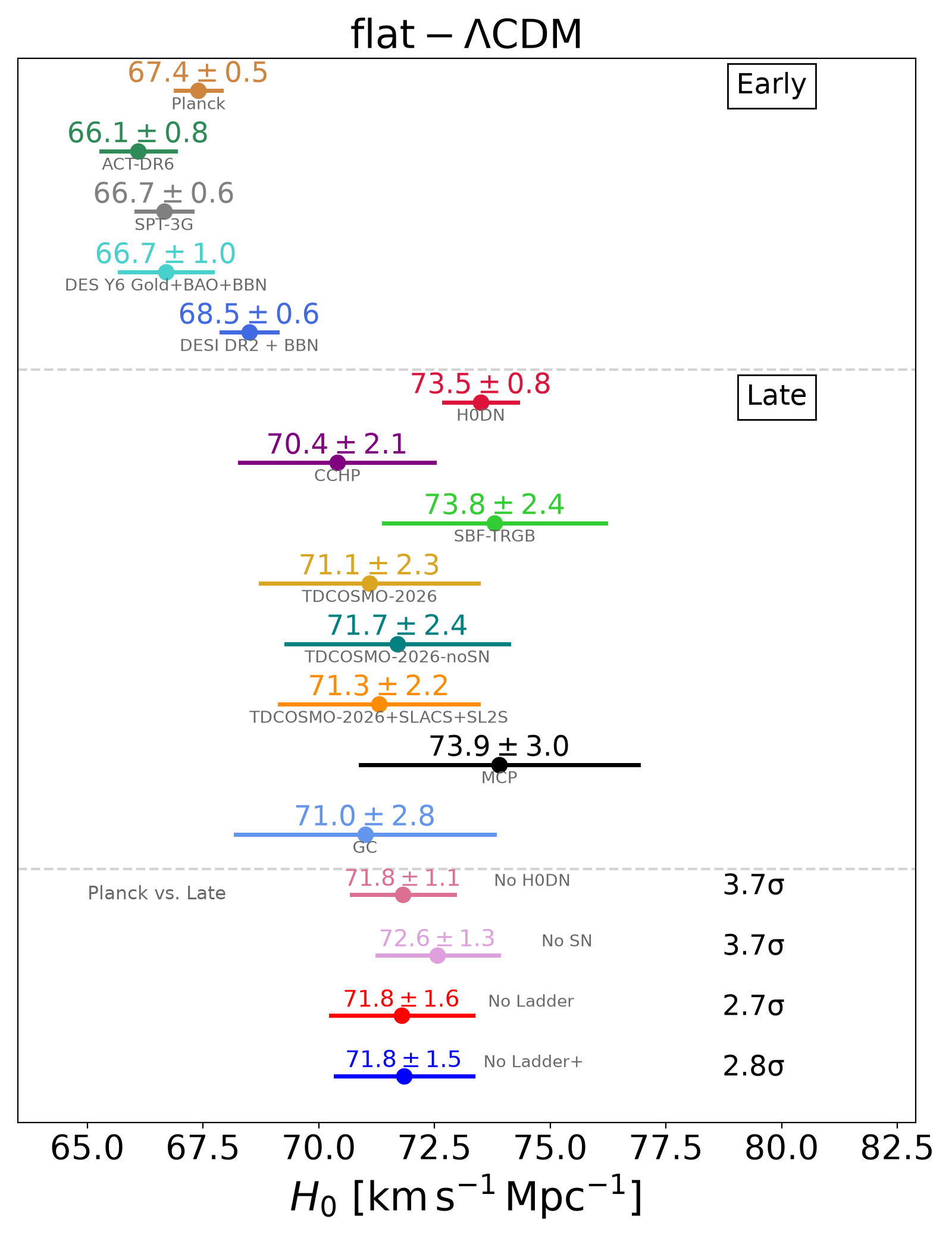}
    \caption{State of the Hubble tension in flat $\Lambda$CDM. The top panel shows early universe probes from \textit{Planck} \citep{Planck2020}, ACT-DR6 \citep{ACTDR6}, SPT-3G \citep{SPT26}, DES-YR6-Gold+BAO+BBN \citep{DES-YR6}, and DESI-DR2+BBN \citep{DESIDR2}. The middle panel shows late universe measurements that have precision 5\% or better. These include: our own (TDCOSMO-2026); the local distance network based on a combination of methods \cite[][H0DN]{H0DN:2026}; the local distance ladder based on TRGB and SN Ia by \citet[][CCHP]{Freedman24}; surface brightness fluctuations calibrated by TRGB \citep{Jensen2025}, megamasers \citep[][MCP]{Pesce:2020},  and globular clusters \citep[][GC]{Jimenez:2019}. The bottom panel shows combinations of late universe probes, as described in the text. The combination "No Ladder+" uses TDCOSMO-2026+SLACS+SL2S instead of TDCOSMO-2026. Even excluding the local distance ladder method completely, the combination of TDCOSMO-2026 with megamasers and Globular Clusters (GC) is in 2.7$\sigma$ tension with \textit{Planck}. \label{fig:tension}}
\end{figure*}

For this exercise, we consider six late universe measurements. For conciseness, we limit our comparison to individual measurements that have a precision of 5\% or better, referring the reader to \citet{cosmoverse25} for a complete list of measurements. Based on this cut, \fref{fig:tension} shows measurements using: the local distance network based on a variety of methods \citep{H0DN:2026}; Sn Ia calibrated with the Tip of the Red Giant Branch method \citep[TRGB,][]{Freedman24}; time-delay lenses (TDCOSMO-2026, this work); geometric distances to megamasers \citep{Pesce:2020}; surface brightness fluctuations (SBF) calibrated with the TRGB method \citep{Jensen2025}; and the ages of globular clusters \citep{Jimenez:2019}. The figure also shows five independent early-universe probes to emphasize that the tension is not only with \textit{Planck} \citep{Planck2020} but is more general \cite[see, e.g.,][for a more comprehensive summary and discussion]{cosmoverse25}.

We then construct four combinations. Our goal is not to carry out an exhaustive set of all possible combinations, but to explore whether unknown systematics in any individual measurement or collection of measurements could be driving the tension. We show the combinations in the bottom panel of \fref{fig:tension}. 

The most precise measurement is the one by the H0DN team based on a modern version of the traditional local distance ladder, namely a covariant combination of a large variety of literature measurements into a consistent direct measurement  \citep{H0DN:2026}. That measurement itself is in more than 6$\sigma$ tension with \textit{Planck}.  As an alternative to H0DN and combinations explored therein, we  construct four combinations: i) excluding H0DN and CCHP and  not using supernova priors on $\Omega_{\rm m}$ for the TDCOSMO ("No SN" for simplicity); excluding H0DN ("No H0DN"); excluding the local distance ladder altogether ("No Ladder" and "No Ladder+"). The combination labeled "No Ladder+" uses the TDCOSMO-2026+SLACS+SL2S measurement instead of the TDCOSMO-2026 baseline. The tension with \textit{Planck} is 2.7$\sigma$ or greater.

\section{Conclusions and future outlook }
\label{sec:conclusions}

We have presented the measurement of the Hubble constant $H_0$ in flat $\Lambda$CDM based on the blind analysis of 11 multiply-imaged quasars with measured time delays. The main results can be summarized as follows:

\begin{enumerate}
    \item We find $H_0=\hnotfiducial$\ksmpc in our baseline analysis. With the addition of constraints from the non-time delay SLACS and SL2S samples we find $H_0=71.3^{+2.2}_{-2.1}$\ksmpc. The result is robust to different analysis choices. Systematic errors are at the percent level, subdominant with respect to the statistical uncertainties.
    \item Our precision has now reached 3\%, a substantial improvement with respect to the 5\% precision reached in our 2025 milestone \citep{TDCOSMO2025}. 
    \item Among late-universe probes, our precision is second only to the local distance ladder, and our measurement is completely independent of it.
    \item By combining our results with a number of independent late universe probes, we find that the tension with \textit{Planck} is significant at the 2.7$\sigma$ level or more, even excluding the traditional local distance ladder altogether. 
\end{enumerate}

While our results are a major step forward with respect to previous work, our precision is not quite sufficient to test the Hubble tension independently of all other methods.
Further improving the precision will require expanding the sample size and reducing the uncertainties per system.

Our collaboration is currently working on both fronts. Two approved JWST programs (JWST-GO-7184, PI: Millon; JWST-GO-9637, PI: Treu) are expanding the number of quadruply imaged quasars with NIRCam imaging from 3 to 19 and the number with NIRSpec spectroscopy from 6 to 18. These samples are predicted to enable a measurement of $H_0$ with 1.5\% precision in flat $\Lambda$CDM. 

Beyond that, the large number of lensed quasars and supernovae expected to be discovered by \textit{Euclid}, Rubin, and Roman will enable the determination of $H_0$ with 1-2\%  precision in general cosmologies \citep{Erickson26,Shajib25lsst}. 

\section{Author contributions and Acknowledgments}
\paragraph{Acknowledgments:} We thank Richard Anderson and Lloyd Knox for comments that improved the manuscript. MM and HK acknowledge support by the Swiss National Science Foundation (SNSF) through Ambizione grant PZ00P2\_223738. AG acknowledges support by the SNSF through PostDoc Mobility Return grant P5R5-2\_235360. The UCLA group acknowledges support by NSF through grants NSF-AST-2407277, by NASA through grant JWST-GO-2974, and from the Moore Foundation through grant 8548.AJS acknowledges support by NASA through grant JWST-GO-2974.KCW is supported by JSPS KAKENHI Grant Numbers JP24K07089, JP24H00221.E.P. is supported by JSPS KAKENHI Grant Number JP24H00221. DS acknowledges the support of the Fonds de la Recherche Scientifique-FNRS, Belgium, under grant No. 4.4503.1. CDF acknowledges support for this work from the US National Science Foundation under Grant No. AST-2407278. DJ acknowledges support by the First Rand Foundation, South Africa, and the Centre National de la Recherche Scientifique of France. PF acknowledges support from the French Agence Nationale de la Recherche through the ELROND project (ANR-23-CE31-0002). This work was supported by the Agencia Estatal de Investigación (AEI), Ministerio de Ciencia, Innovación y Universidades, Spain, under the project PID2024‑155455NB‑I00, within the 2024 Call for Knowledge Generation Projects. FC has also received support through ERC Synergy RedH0T under grant agreement 101225120. V.M. acknowledges support from ANID FONDECYT Regular grant number 1231418, and Centro de Astrof\'{\i}sica de Valpara\'{\i}so CIDI 21. SB and XH are supported by DoE Grant DE-SC0026113. The research reported here is supported by Stanford University under Contract No. DE-AC02-76SF00515 with the U.S. Department of Energy, Office of Science, Office of High Energy and Nuclear Physics. T.M. acknowledge support by JSPS through KAKENHI Grant Number 25K24918. MS acknowledge support under NASA grant 80NSSC22K1294.
\paragraph{Author contributions:} See the companion TDC26 paper.

\bibliographystyle{unsrtnat} 
\bibliography{references}

\end{document}

%% file: packages.tex
\usepackage[english]{babel}

\usepackage[utf8]{inputenc}
\usepackage[T1]{fontenc}
\usepackage{ucs}
\usepackage{caption}
\usepackage{microtype}
\usepackage{newtxtext,newtxmath}
\usepackage{color}
\usepackage{lettrine}
\usepackage{gensymb}
\usepackage{tablefootnote} 
\usepackage{threeparttable}

\usepackage{url}
\usepackage{hyperref}

\usepackage{xspace}
\usepackage{lastpage}
\usepackage{stmaryrd}

\usepackage{graphicx}
\usepackage{empheq}
\usepackage{ifthen}
\usepackage{tensor}
\usepackage{paralist}
\usepackage{float}

\usepackage{slashed}
\usepackage{amsmath}
\usepackage{amsfonts}

\usepackage{amssymb}
\usepackage{mathrsfs}
\usepackage{extarrows}
\usepackage{verbatim}
\usepackage{upgreek}
\usepackage{wasysym}
\usepackage{multirow}

\usepackage{natbib}
\usepackage{makecell}
\usepackage{tabularx}